\documentstyle[preprint,aps]{revtex}
\begin{document}
\newcommand{\bfsigma}{\mbox{\boldmath $\sigma$}}
\draft
\title{SMMC method for
two-neutrino double beta
decay}

\author{P.B. Radha${}^1$, D. J. Dean${}^{1,2}$, S. E. Koonin${}^1$, T. T. S.
Kuo${}^3$, K. Langanke${}^1$,\\
A. Poves${}^{1,4}$,
J. Retamosa${}^4$, and P. Vogel${}^5$}
\address{${}^1$W. K. Kellogg Radiation Laboratory, California
Institute of Technology\\ Pasadena, California 91125 USA\\
${}^2$ Physics Division, Oak Ridge National Laboratory, P.O. Box 2008,
\\Oak Ridge, Tennessee 37381 USA\\
${}^3$ Physics Department, SUNY Stony Brook, New York 11794 USA\\
${}^4$ Departmento de F\'isica Te\'orica C-XI\\
Universidad Aut\'onoma de Madrid,
E-28049 Madrid, Spain\\
${}^5$ Division of Physics, Mathematics and Astronomy\\
California Institute of Technology,
Pasadena, California 91125 USA}
\date{\today}
\maketitle

\begin{abstract}
Shell Model Monte Carlo (SMMC) techniques are used to
calculate two-neutrino double beta decay matrix elements.
We validate the approach against direct diagonalization  for
$^{48}$Ca in the complete $pf$-shell using the KB3 interaction.
The method is then applied to
the decay of $^{76}$Ge in the
$(0f_{5/2},1p,0g_{9/2})$ model space using a newly calculated realistic
interaction. Our result
for the matrix element is $0.13\pm0.05$ MeV$^{-1}$, in
agreement with the experimental value.

\end{abstract}


\narrowtext

The double beta ($\beta\beta$) decay of a nucleus is a
rare second order weak process
\cite{haxton,moe}.
The as yet unobserved neutrinoless mode is of fundamental interest,
as it would signal a
neutrino mass, lepton number non-conservation, or
admixtures of right handed weak currents.
In contrast, the existence of the $2\nu$ mode has been firmly
established
(see the review in Ref \cite{moe}).
The ability to accurately describe this latter process is an
important element in the interpretation of limits on neutrinoless
decays. Unfortunately, it seems that $2\nu$ matrix elements are
highly suppressed and so depend sensitively on small, poorly
determined parts of the nuclear wavefunctions.

Most recent
calculations of $2\nu\beta\beta$ matrix elements for nuclei heavier than
$^{48}$Ca rely on the quasi-particle random phase approximation (QRPA)
\cite{moe}.
While this approach
is computationally simple and includes
many features of the two-body interaction known to be
relevant for
$\beta\beta$ decay, the
calculated matrix elements are uncertain
because of their great sensitivity to the
$J=1^+, T=0$ particle-particle interaction
\cite{zirnbauer}.
The interacting shell model offers a more
microscopic approach to the problem.
Complete $0\hbar\omega$ shell model calculations \cite{sdshell sn100}
not only recover more quenching of
Gamow-Teller (GT) transitions than
QRPA calculations, but also are in agreement with observations
(after the universal renormalization of $g_A$ to $1.0$).
However, computational limitations have restricted
shell model calculations
of the $2\nu \beta\beta$ decay matrix element to $^{48}$Ca,
the lightest of all $\beta\beta$ candidates.

In this Letter we show how SMMC methods can be used to calculate
$2\nu \beta \beta$ decay matrix elements.
We first calculate
the decay of $^{48}$Ca in the
complete $pf$-shell and validate our method against
direct diagonalization. We then present results for the decay of
$^{76}$Ge, one of the few nuclei where the $2\nu\beta\beta$ decay
has been precisely measured and where the best limits on the
$0\nu$ decay have been obtained \cite{dbdge1}. Our
calculation in the $(0f_{5/2},1p,0g_{9/2})$ orbitals,
which is impractical using traditional shell model
methods, is the first for $2\nu\beta\beta$ decay in such a large
model space.

The $2\nu\beta\beta$ matrix element between the $0^{+}$ ground states of
the initial and final even-even nuclei is given by \cite{Petr's book},
\begin{equation}
M^{2\nu}=\sum_m {\langle 0_f^+| {\bf G} |1^+_m \rangle \cdot \langle
1^+_m| {\bf G} |0_i^+\rangle \over E_m-(E_i^0+E_f^0)/2}.
\end{equation}
Here, $|0_i^+\rangle$ $(|0_f^+\rangle)$ is the ground state
of the initial (final)
nucleus with energy $E_i^0$ $(E_f^0)$; $|1^+_m\rangle$ are the $1^+$
states of the intermediate odd-odd nucleus with energies $E_m$; and
{$\bf G$} is the GT operator,
$\sum_l \bfsigma_l\tau^-_l$, where
$\bfsigma_l$ is the Pauli spin operator
for nucleon $\it l$ and
$\tau^-_l$ is the isospin lowering operator that changes a neutron
into a proton.

Previous shell model calculations for nuclei heavier than $^{48}$Ca
(\cite{haxton} and references therein)
have invoked the so-called closure approximation, where
the matrix element is written as
$M^{2\nu}=M_c/\bar E$,
with
$M_c=\langle 0_f^+| {\bf G} \cdot {\bf G} |0_i^+ \rangle$
and $\bar E$ an average energy denominator.
As there is no prescription for choosing the average energy denominator
(and even the closure matrix element is usually calculated in a severely
truncated basis),
the uncertainty in this
approximation is difficult to estimate.

To calculate the exact $2\nu\beta\beta$ matrix element,
Eq. (1), we consider the function
\begin{equation}
\phi(\tau,\tau^{\prime})={{\rm Tr}[e^{-(\beta-\tau-\tau^{\prime}) H}
{\bf G^\dagger}
\cdot {\bf G^\dagger} e^{-\tau H} {\bf G} e^{-\tau^{\prime} H} \cdot {\bf G}]
\over {\rm Tr}[e^{-\beta H}]} ,
\end{equation}
where $H$ is the many-body Hamiltonian and the trace is
over all states of the initial nucleus.
The quantities $(\beta-\tau-\tau^{\prime})$ and
$\tau$ play the role of the inverse temperature in the parent and
daughter nucleus respectively. A spectral expansion of $\phi$ shows that large
values of these parameters guarantee cooling to the parent and
daughter ground states. In these limits, we note that
$\phi(\tau,\tau^{\prime}=0)$
approaches $e^{-\tau Q} |M_c|^2$,  where $Q=E_i^0-E_f^0$
is the energy release, so that a calculation of $\phi(\tau,0)$
leads directly to the closure matrix
element. If we then define
\begin{equation}
\eta(T,\tau)\equiv \int_0^T d\tau^{\prime} \phi(\tau,\tau^{\prime})
e^{-\tau^{\prime} Q/2},
\end{equation}
and
\begin{equation}
M^{2\nu}(T,\tau)\equiv{\eta(T,\tau) M^{*}_{c} \over \phi(\tau,0)},
\end{equation}
it is easy to see that in the limit of large
$\tau$, $(\beta-\tau-\tau^{\prime})$, and $T$,
$M^{2\nu}(T,\tau)$ becomes
independent of these parameters and is equal to the matrix element
in Eq. (1).

We use SMMC methods \cite{lang} to calculate $\phi(\tau,\tau^{\prime})$,
and hence $M^{2\nu}$. These techniques scale more gently than
direct diagonalization with
the number of valence nucleons and single particle orbits and so
allow calculations larger then possible otherwise.
They are based on
the discretization of the many-body propagator, $e^{-\beta H}$,
into a finite number of ``time'' slices, $N_t$, each of duration
$\Delta\beta=\beta/N_t$. At each time slice the-many body propagator
is linearized via the Hubbard-Stratonovich transformation \cite{hs};
observables are then calculated as
expectation values in the canonical ensemble of nuclear states.

To circumvent the ``sign problem'' encountered in the SMMC
calculations with realistic interactions,
we use the extrapolation procedure outlined in \cite{thermal}.
One defines a set of Hamiltonians
$H(g,\chi)= (1-{(1-g)\over \chi}) H_G+g H_B$
such that $H(g=1,\chi)=H$ is the physical Hamiltonian and
$H_{G,B}$ are the ``good'' and ``bad'' parts of the Hamiltonian,
respectively.
For $g\leq0$, $H(g,\chi)$ is free of the sign problem; the matrix
elements are therefore calculated for several values $g\leq0$ and
extrapolated to $g=1$.
The value of
$\chi$ is chosen to make the linear $g$-extrapolation as smooth as
possible.

To validate our method, we calculated the matrix elements for $^{48}$Ca
in the complete $pf$-shell with the KB3 interaction \cite {kb3}
for six equally spaced $g$ values between
-1.0 and 0.0 using $\chi=4$ and extrapolated to the physical result at $g=1$.
Each calculation involved 2500-3500 Monte Carlo samples and
was performed at $\beta=2$ MeV$^{-1}$
with $N_t=48$.
The direct diagonalization
calculations for $^{48}$Ca with which we compare our results were performed
using an implementation of the Lanczos
algorithm \cite{lanczos}.  We calculated both the closure
and exact matrix elements for the same Hamiltonians, $H(g,\chi)$,
as used in the SMMC.

We found the slope of $\ln[\phi(\tau,0)]$ to be in good
agreement with that expected from the difference of the energies
for $^{48}$Ti and $^{48}$Ca (Fig. 1) and extracted $|M_c|$ from the
intercept.
The SMMC closure matrix elements for $g\leq0$
are in very good
agreement with the direct diagonalization results (Fig. 1)
indicating that our temperatures are sufficiently low to
correctly calculate the closure matrix element from the ground
state of $^{48}$Ca to $^{48}$Ti. However, the
direct diagonalization
calculations show a small curvature near $g=1.0$ that the extrapolation
cannot reproduce. Our linear extrapolation of the closure matrix
element, which takes place over
almost a factor of $20$,
therefore
underestimates the physical ($g=1.0$) calculation. We obtain $-0.21\pm0.29$
for the closure matrix element to be compared with the
direct diagonalization result of $0.29$. As the natural scale for
$M_c$ is given by the sum rule \cite{moe} as
$\approx 21$, we may conclude that the SMMC
successfully reproduces
the shell model suppression of a factor of 70.

The calculation of the function $\phi(\tau,\tau^{\prime})$ was performed
for $\tau=0.5$ MeV$^{-1}$ and for thirteen $\tau^{\prime}$ values spaced
equally between 0.0 and 0.5 MeV$^{-1}$. This combination of parameters was
checked to give converged results for the matrix element.
We then calculated $\eta(T,\tau)$ for thirteen values of
$T\leq0.5$ MeV$^{-1}$; the upper limit of $T$ is sufficiently large for the
integral in Eq. (3) to converge. From these, we obtained
$M^{2\nu}(T,\tau)$ (Eq. (6)) as shown in Fig. 2
for some representative values of $g$
\cite{footnote}.
In Fig. 2, we show the good agreement between the
SMMC matrix elements and direct
diagonalization for $g\leq0$. Even though the value of $\chi=4.0$
was chosen to make the linear extrapolation as smooth as possible,
the direct diagonalization results still have a small curvature.
For the exact $2\nu$ matrix element we obtain an extrapolated
value of  $0.15 \pm 0.07$ MeV$^{-1}$
whereas
the calculation of Caurier {\it et al.}
\cite{caurier}
(including the erratum in \cite{kb3}) gives $0.08$ MeV$^{-1}$.
There is thus agreement within the uncertainty.

We now apply the SMMC method to a heavier nucleus,
where direct diagonalization is
not possible. In particular,
we calculate the $2\nu$ matrix element for $^{76}$Ge
using an effective interaction based on the Paris potential in the
$(0f_{5/2},1p,0g_{9/2})$ orbitals, with the single
particle energies taken from the levels of $^{57}$Ni relative to
the $^{56}$Ni core \cite{mottelson}.
This interaction has been constructed
using a {\it {G}}-matrix folded-diagram method, in close analogy
with the calculations carried out by Shurpin {\it et al.}
\cite{sks83} and by
Dean {\it et al.} \cite{sdshell sn100}.
The model space comprises
some $10^{8}$ configurations, so that our SMMC calculation is
significantly larger than previous
shell model treatments of $^{76}$Ge \cite{haxton}. While it
avoids spurious excitations of the center of mass, it
does not include all spin-orbit pairs of orbitals and thus
does not obey the Ikeda sum rule for GT strengths.
However, this model space (with the choice
of an appropriate effective interaction)
should adequately describe those low-lying states
expected to be the most important  for $2\nu\beta\beta$ decay
\cite{ericson}.

We performed the $^{76}$Ge calculation at $\beta=2.5$ MeV$^{-1}$
with $N_t=60$.
The effective interaction used reproduces the experimental
mass splitting of $^{76}$Ge and $^{76}$Se very well:
$20.41 \pm 1.3$ MeV
compared to the experimental splitting of $20.72$ MeV
(the Coulomb energy was calculated following Ref  \cite{myers}).
Our value for the $\beta^-$ strength of $^{76}$Ge is
$B(GT_-)=\langle {\bf G^\dagger} \cdot {\bf G} \rangle =
12.6\pm0.3$ and we find an energy centroid of
$6.3\pm0.2$ MeV,
while the experimental values are $19.9$ and
$9.1$ MeV respectively \cite{geexp}; the discrepancies are almost
certainly due to
not having the $g_{7/2}$ and $f_{7/2}$ orbitals in our model space.
The missing
high-energy strength should not affect the low-lying states of $^{76}$As
that are important for $2\nu\beta\beta$ decay.
We find a vanishingly small $\beta^+$ strength for $^{76}$Se, $B(GT_+)=
-0.05\pm0.03$.
This strength is identically zero in the independent particle model
and it is generated only by the smearing of the fermi surface
due to the interaction.
As our $g$-extrapolation takes
place over a range of pairing strengths, $B(GT_+)$ varies by
almost an order of magnitude between $g=-1.0$ and $g=1.0$, leading to
large uncertainties in the physical value.
A detailed study of this mass region with our
effective interaction is planned.

We performed two independent sets of calculations for both the
closure and the
exact matrix element using the $\chi=4$ and $\chi=\infty$ families
of $H(g,\chi)$. The extrapolations to $g=1.0$ were linear
for both the closure and the exact matrix elements in both cases.
Our results for the closure matrix elements are $-0.36\pm0.34$ and
$0.08\pm0.17$
for $\chi=4$ and $\chi=\infty$ respectively. These are to be compared
with the truncated shell model calculation of Haxton {\it et al.}
\cite{haxton} (using a different
effective interaction) that
resulted in a value of $2.56$.

We find consistent exact matrix elements for the $\chi=4$ and $\chi=\infty$
cases (Fig. 3). Our results are $0.12\pm0.07$ MeV$^{-1}$ and
$0.14\pm0.08$ MeV$^{-1}$ respectively (a combined value of $0.13\pm0.05$
 MeV$^{-1}$),
while the experimental value of this matrix element (using $g_A=1.26$)
is $0.14\pm0.01$ MeV$^{-1}$ \cite{dbdge1}. However, shell model calculations of
ordinary $\beta$-decay consistently suggest that $g_A$ is
renormalized to $1.0$ in the nuclear medium \cite{sdshell sn100},
in which case the experimental matrix element is $0.22\pm0.01$ MeV$^{-1}$.

There has been no previous
shell model calculation of $M^{2\nu}$. Haxton {\it et al.}
\cite{haxton} obtained an estimate in the closure approximation by
taking the average energy denominator to be the position of the $\beta^-$
GT resonance in $^{76}$Ge ($9.4$ MeV).
We find significantly smaller
values of $\bar E \equiv M_c/M^{2\nu}$
($-3.0\pm3.3$ MeV and $0.57\pm1.26$ MeV for $\chi=4$
and $\infty$, respectively), evidently due to large cancellations
among the various terms in (1).

In this Letter, we have demonstrated an SMMC method to calculate
$2\nu\beta\beta$ decay matrix elements. Results
for the $^{48}$Ca decay compare
well with those from direct diagonalization.
We have also
calculated the matrix element for $^{76}$Ge in a model space
significantly larger than previous calculations, and
obtain a result that is in reasonable
agreement with experiment.
A more detailed description of
these calculations will be given elsewhere, and work is in progress to
calculate the matrix elements of several other, heavier nuclei.

This work was carried out under grants from the NSF and the DOE.
Computational cycles were provided by the Concurrent Supercomputing
Consortium on the Intel Touchstone Delta and the Intel Paragon and
on the IBM SP2 at the Maui High Performance Computing Center.

\begin{figure}
\caption{Upper: $\ln[\phi(\tau,0)]$
for $^{48}$Ca calculated at $\beta=2.0$ MeV$^{-1}$ with
$N_t=48$. The lines are best fits.
Lower:
SMMC and direct diagonalization
closure matrix elements for $^{48}$Ca.
The SMMC points are
linearly extrapolated to $g=1.0$.}
\label{fig1}
\end{figure}
\begin{figure}
\caption{Upper: $M^{2\nu}(T,\tau)$  for $^{48}$Ca
calculated at $\beta=2.0$ MeV$^{-1}$ with $N_t=48$.
The points at large $T$ show
the asymptotic value
$(T\rightarrow \infty$) of the matrix elements;
lines are drawn to guide the eye.
Lower: SMMC exact matrix
elements and the direct diagonalization results for $^{48}$Ca.
The SMMC matrix elements are linearly extrapolated
to $g=1.0$.}
\label{fig2}
\end{figure}
\begin{figure}
\caption{SMMC exact matrix elements for $^{76}$Ge calculated
using the Hamiltonians $H(g,\chi)$ with $\chi=4$
and $\chi=\infty$. The lines are linear fits to
the points in both cases. The extrapolated values and the experimental
result of Ref. [5] are shown staggered around $g=1.0$
for clarity.}
\label{fig3}
\end{figure}

\begin{references}
\bibitem{footnote}To determine the
asymptotic ($T\rightarrow\infty$) value of $M^{2\nu}(\tau,T)$, we
construct an ensemble of Gaussian-distributed $\phi(\tau,\tau^\prime)$
with mean and standard deviations as calculated in SMMC.
Each $M^{2\nu}(\tau,T)$ is then fit to the form $a-be^{-cT}$
between $T=0.17$ MeV$^{-1}$ and $T=0.5$ MeV$^{-1}$,
and the average and standard deviation
of $a$ gives the required matrix element and its error
(shown at large $T$ in the upper panel of Fig. 2).
\end{references}
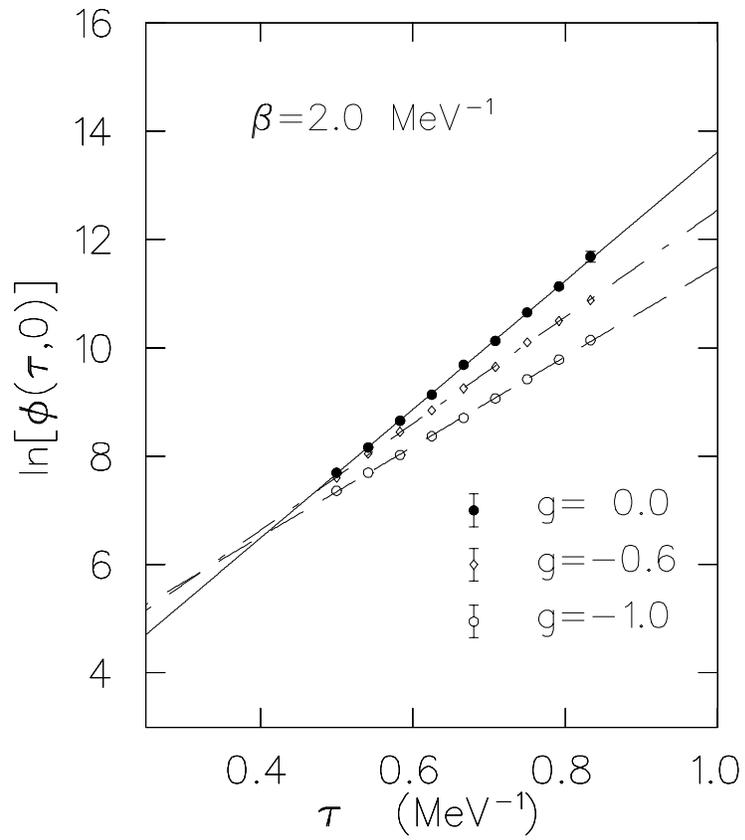
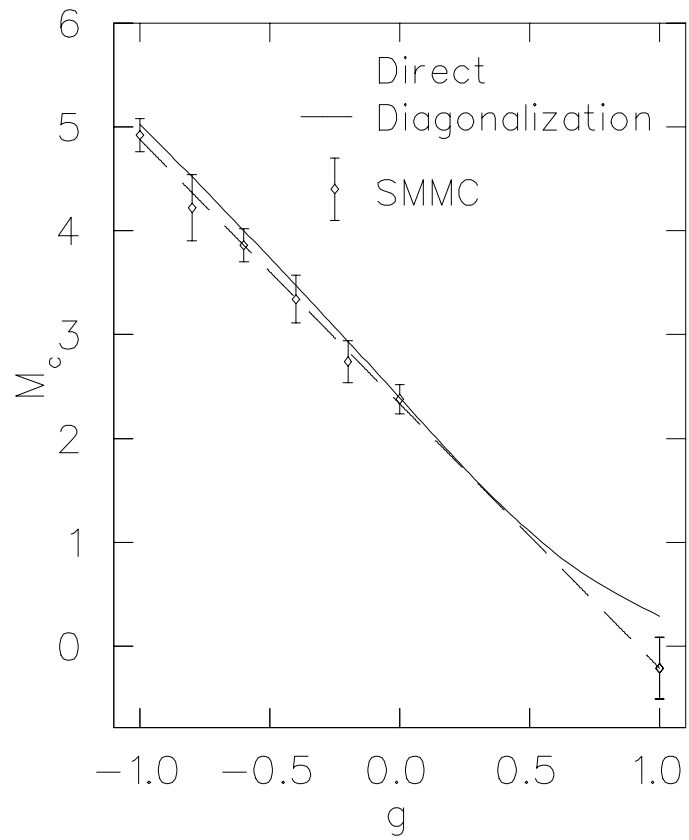
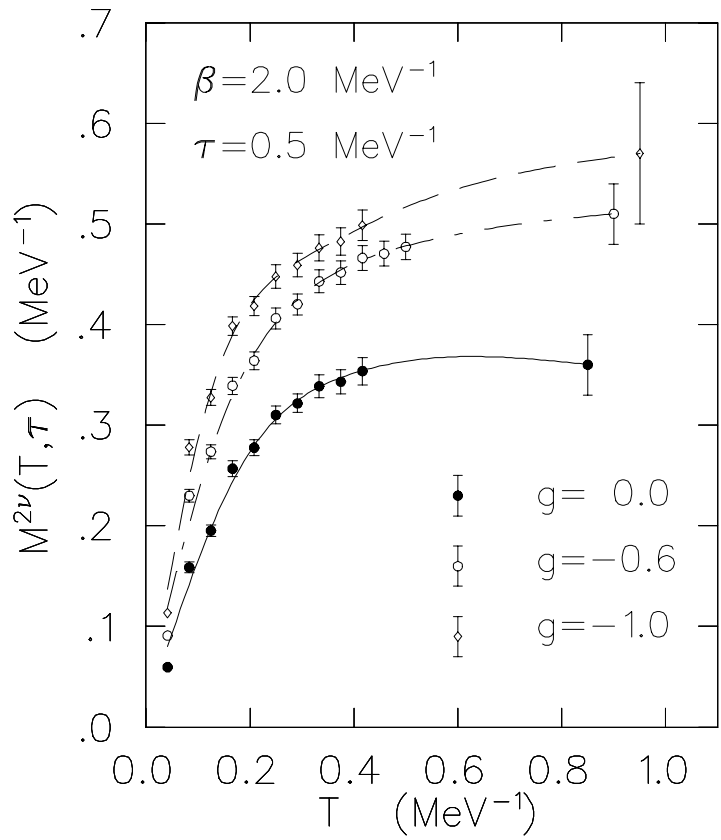
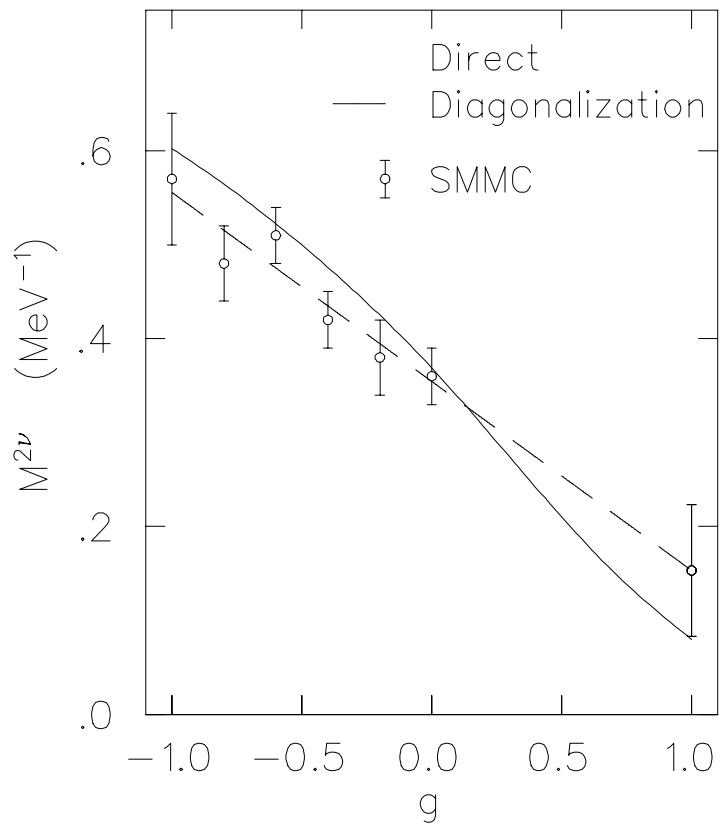
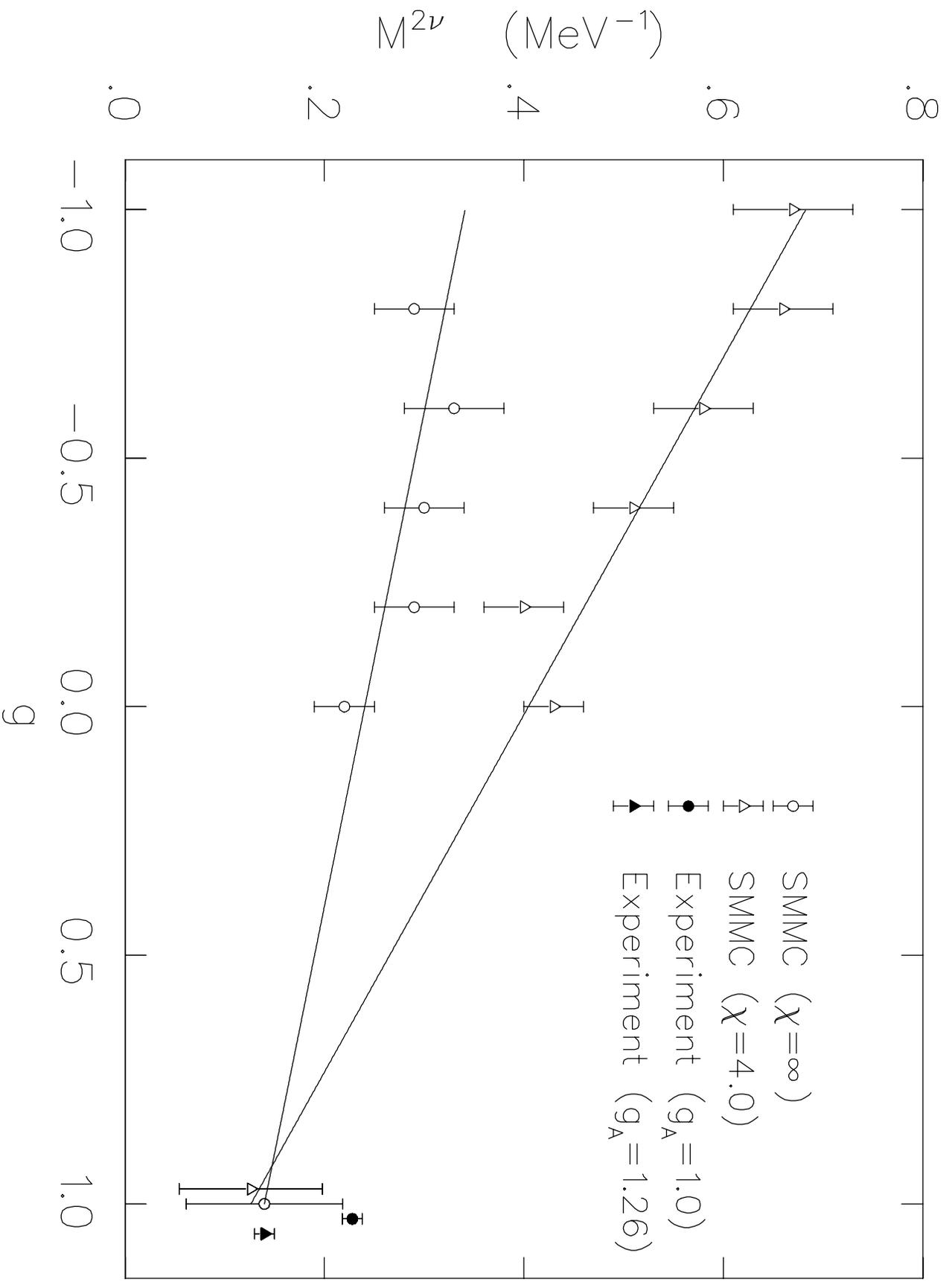
\end{document}